# Synthesis, structure and magnetic properties of honeycomb-layered Li$_3$Co$_2$SbO$_6$ with new data on its sodium precursor, Na$_3$Co$_2$SbO$_6$

M.I. Stratan[a], I.L. Shukaev[b], T.M. Vasilchikova[a], A.N. Vasiliev[a,c,d], A.N. Korshunov[e,f], A.I. Kurbakov[e,f,] V.B. Nalbandyan[b] and E.A. Zvereva[a]*

[a] *Faculty of Physics, Moscow State University, Moscow 119991, Russia*
[b] *Southern Federal University, 7 Zorge street, Rostov-on-Don, 344090, Russia*
[c] *National Research South Ural State University, Chelyabinsk 454080, Russia*
[d] *National University of Science and Technology "MISiS", Moscow 119049, Russia*
[e] *Petersburg Nuclear Physics Institute – NRC Kurchatov Institute, Gatchina 188300, Russia*
[f] *Faculty of Physics, St.Petersburg State University, St. Petersburg 198504, Russia*
[*]*E-mail: zvereva@mig.phys.msu.ru*
CCDC 1883837 for Li$_3$Co$_2$SbO$_6$

Li$_3$Co$_2$SbO$_6$ is prepared by molten salt ion exchange and its structure refined by the Rietveld method confirming the honeycomb-type Co/Sb ordering of its Na precursor. Monoclinic rather than trigonal symmetry of Na$_3$Co$_2$SbO$_6$ is directly demonstrated for the first time by peak splitting in the high-resolution synchrotron XRD pattern. The long-range antiferromagnetic order is established at $T_N \approx 6.7$ K and 9.9 K in Na$_3$Co$_2$SbO$_6$ and Li$_3$Co$_2$SbO$_6$, respectively, confirmed by both the magnetic susceptibility and specific heat. Spin-wave analysis of specific heat data indicates the presence of 3D AFM magnons in Na$_3$Co$_2$SbO$_6$ and 2D AFM magnons in Li$_3$Co$_2$SbO$_6$. The field dependence of the magnetization almost reaches a saturation in moderate magnetic fields up to 9 T and demonstrates characteristic features of magnetic field induced spin-reorientation transitions for both A$_3$Co$_2$SbO$_6$ (A = Na, Li). Overall thermodynamic studies show that the magnetic properties of both compounds are very sensitive to external magnetic field, thus predicting non-trivial ground state with rich magnetic phase diagram. The ground state spin configuration of Li$_3$Co$_2$SbO$_6$ has been determined by low-temperature neutron powder diffraction. It represents a ferromagnetic arrangement of moments in the honeycomb layers with antiferromagnetic coupling between adjacent layers.

## 1. Introduction

Layered mixed oxides containing cations with partially filled d shell are interested for both electrochemical energy storage technologies and fundamental solid state physics. Of this vast field, we only shall discuss here honeycomb-ordered A$_3$M$_2$SbO$_6$ antimonates where A and M are a univalent metal and a divalent transition metal (Co, Ni, Cu), respectively. First studied structurally,[1-4] this family later attracted attention as possible electrode materials for Li-ion[5-7] or Na-ion[8-16] batteries (with M=Ni[5-15] and Cu[16]) and as a playground for non-trivial physical phenomena, with M=Cu,[17-26] Ni,[26-31] and Co.[30-33] The honeycomb lattice can be split into two identical sub-lattices. In contrast to the triangular lattice, the honeycomb lattice is not geometrically frustrated for the antiferromagnetic interactions but it has the minimum coordination number N=3 possible for any regular 2D lattice. Thus the quantum fluctuations are expected to be weaker than in the one-dimensional (1D) chain case (N=2), but stronger than those in other two-dimensional lattices (for example, N=4 in case of square lattice). Hence, the antiferromagnetic order for the honeycomb lattice is more fragile. As expected for pure 2D honeycomb lattice, it has no magnetic order at any finite temperature. Switching on the interlayer coupling and anisotropies and their influence on the ground state in such systems are largely unexplored at present.

$A_3Ni_2SbO_6$ (A = Na, Li) and $Na_3Co_2SbO_6$ antimonates are superlattices of α-$NaFeO_2$ (or $LiCoO_2$) type layered structures with honeycomb ordering of $Ni^{2+}$ or $Co^{2+}$ and $Sb^{5+}$ on $Fe^{3+}$ sites. For $Na_3Co_2SbO_6$, first studied by some of the present authors,[2] basic magnetic properties and low-temperature magnetic structure were reported.[32,33] For $Li_3Co_2SbO_6$, however, no data on crystal structure or properties have been published yet, besides a conference paper[34]. It could only be prepared by ion exchange[35] from the Na precursor because direct high-temperature synthesis yielded a different, orthorhombic, polymorph[35] isostructural with $Li_3Ni_2TaO_6$.[36]

We report here details of preparation, crystal structure refinement and magnetic properties of $Li_3Co_2SbO_6$ in comparison with its sodium precursor $Na_3Co_2SbO_6$.

## 2. Experimental

### 2.1. Sample preparation

$Na_3Co_2SbO_6$ was prepared by solid-state synthesis as reported previously.[37,38] It was converted to $Li_3Co_2SbO_6$ by ion exchange in molten salts. Preliminary experiments showed that treatment in $LiNO_3$ at 300 °C yielded black products with unit cell volumes considerably smaller than that for $Li_3Ni_2SbO_6$[27] although $Co^{2+}$ is considerably larger than $Ni^{2+}$.[39] This indicated oxidation to $Co^{3+}$ and necessity of selecting non-oxidizing salts. Therefore, ion exchange was performed in a low-melting mixture of LiCl and LiBr, taken in a 15-fold excess, at 540 °C for 45 min. Initially, to avoid hydrolysis, the excess salts were washed out with anhydrous methanol;[35] in this work, however, essentially the same product was obtained after washing with weak aqueous solution of $Li_2CO_3$ (~0.3 wt. %) and drying at 120 °C. The colour of the product was reddish-brown, similar to that of the Na precursor. The large sample for neutron diffraction was prepared separately from the small sample used for X-ray diffraction, magnetic and heat capacity studies and this explains slight difference in the temperatures of phase transition.

### 2.2. X-ray diffraction studies

Crystal structure of $Li_3Co_2SbO_6$ was refined using the GSAS+EXPGUI suite[40,41] and XRD data measured previously[35] with the Rigaku rotating anode instrument equipped with a secondary beam graphite monochromator. The starting $Na_3Co_2SbO_6$ sample was studied with better resolution at the ID31 line of ESRF (now the diffractometer is moved to the ID22 position) with wavelength of 0.39985Å.

### 2.3. Magnetization and specific heat measurements

The temperature dependences of the magnetic susceptibility were measured at the magnetic field $B$ = 0.1T in the temperature range 2 – 300 K by means of a Quantum Design PPMS 9 system. In addition, the isothermal magnetization curves were obtained in external fields up to 9 T at various constant temperatures ($T$ = 1.8 – 9K) after cooling the sample in zero magnetic field.

Specific heat measurements were carried out by a relaxation method using a Quantum Design PPMS system. Data were collected at zero magnetic field in the temperature range 1.8 – 300K.

### 2.4. Neutron powder diffraction

The magnetic structure of the $Na_3Co_2SbO_6$ compound has already been established according to powder neutron diffraction data[32]. Therefore, we investigated with neutrons only the $Li_3Co_2SbO_6$ composition. The low-temperature neutron diffraction experiments were carried out on the cold neutron two-axis diffractometer G4.1 located at the ORPHEE reactor, Laboratory Léon Brillouin (Saclay, France). This diffractometer has a high luminosity in the entire angular range and a high resolution at low diffraction angles, where magnetic neutron scattering is recorded. The wavelength of the incident neutron beam was 2.428 Å. The sample was encased in a vanadium container with an inner radius of 6 mm. Neutron diffraction data were collected in the 2θ range of 6° – 86° with steps of 0.1°. Measurements were carried out at the lowest reached temperature of 1.5 K, then at T = 3, 5, 7, 8, 9, 11, 30 and 80 K. All diffraction patterns were treated with the Rietveld method using the FULLPROF suite.[42,43]

## 3. Results and discussion

### 3.1. Structural data

#### 3.1.1. $Na_3Co_2SbO_6$

Since the aristotype structure of α-$NaFeO_2$ is trigonal (rhombohedral), it was not clear whether Co/Sb ordering leads to symmetry lowering or not. The laboratory XRD pattern of $Na_3Co_2SbO_6$ was indexed with high accuracy on a hexagonal unit cell assuming space group $P3_112$.[2,37,38] On the other hand, its crystal structure was successfully refined from the powder neutron diffraction data within a monoclinic model, space group C2/m,[31,32] although no peak splitting indicating symmetry lowering could be observed. In this work, using high-resolution synchrotron data, peak splitting was observed for the first time (Fig. 1). Thus, the monoclinic model is confirmed most directly. Lattice parameters from the four sources are in good agreement (Table 1); minor discrepancies in cell edges are mostly systematic and may be attributed to the wavelength uncertainties, although slight variations in composition are also possible.[33,38] No re-refinement of the structure was attempted.

### 3.1.2. $Li_3Co_2SbO_6$

In contrast to the pseudo-hexagonal sodium precursor, poor accuracy of the hexagonal indexing, definitely indicated symmetry lowering. The crystal structure was refined within the *C2/m* model.

The stable solution could only be obtained with some restrictions. The gross composition was fixed and all positions were assumed to have full occupations. Sb/Li mixing was forbidden due to large differences in both oxidation states and ionic radii, but Sb/Co and Co/Li mixing were allowed. A set of problems was met due to correlations between occupations, texture parameters, polarization, anisotropic broadening and displacement parameters. Therefore, displacement parameters for the light atoms were fixed to 0.020. Anisotropic displacement parameters were not introduced, even for the heaviest antimony atom, facing to additional correlations. Only accounting for anisotropic broadening and stacking faults parameters (Table 2) allowed sufficient improvement of the quality parameters.

Finally, fraction of Co on Sb site dropped to zero but reasonable thermal parameters and quality criteria could only be obtained with some Li fraction $p$ on Co site. Three $p$ values (0, 0.05 and 0.10) were tested and best results obtained with $p$=0.05 (Table 3). Then, the formula may be written as $(Li_{2.9}Co_{0.1})(Li_{0.1}Co_{1.9}Sb)O_6$. The bond distances and bond valence sums (calculated using the most recent and corroborated set of parameters[44]) are very close to the expected values (Table 4). Due to the small degree of Co/Li mixing (Table 3), it might be ignored in these calculations.

The main results are represented in Figs. 2 and 3 and Tables 2-4. The corresponding crystallographic information file (CIF) has been deposited with the Cambridge Crystallographic Data Centre, Deposition Number 1883837.

### 3.2. Magnetization and specific heat

The static magnetic susceptibility $\chi(T)=M/B$ for $Na_3Co_2SbO_6$ and $Li_3Co_2SbO_6$ is presented in Fig.4. The temperature dependences $\chi(T)$ for both studied compounds demonstrate a sharp maximum with decreasing temperature, which indicates the onset of a long-range antiferromagnetic(AFM) order. The Néel temperatures are estimated to be $T_N$ ~ 6.7K (in consistence with previously reported by Viciu et al.[32]) and $T_N$ ~ 9.8K for sodium and lithium antimonates, respectively. In the high-temperature range $\chi(T)$ obeys a Curie-Weiss law with addition the temperature-independent term $\chi_0$: $\chi=\chi_0+C/(T-\Theta)$, where $C$ is Curie constant and $\Theta$ the Weiss temperature.

The Weiss temperature is found to be negative $\Theta$~-10K for $Na_3Co_2SbO_6$, which indicates the predominance of antiferromagnetic correlations in the paramagnetic phase. On the contrary, for $Li_3Co_2SbO_6$, the Weiss temperature takes positive value $\Theta$ ~ 15 K, which points out the dominant ferromagnetic interactions between cobalt $Co^{2+}$ ions and the existence of competing exchanges in this compound.

For both compounds the $\chi(T)$ dependences deviate noticeably from the Curie-Weiss behaviour at $T<100K$ indicating extended range of short-range correlations typical of low-dimensional (LD) magnets. In the case of the $Na_3Co_2SbO_6$ sample, the $\chi(T)$ deviates to the right side implying the presence of ferromagnetic exchanges, while the $\chi(T)$ for $Li_3Co_2SbO_6$ deviates to the left side relative to Curie-Weiss approximation curve in accordance with the increasing role of antiferromagnetic interactions with temperature decreasing. The effective magnetic moment for $Li_3Co_2SbO_6$ calculated from the obtained Curie constant takes $\mu_{eff} \approx 6.6\mu_B$/f.u. This value is in reasonable agreement with the theoretical estimation $\mu_{theor} \approx 6.5\mu_B$/f.u. assuming that the magnetism in lithium cobalt antimonate is associated with high-spin state $Co^{2+}$ ($S=3/2$) ions with effective g-factor $g=2.3\pm0.1$ (directly obtained from electron spin resonance (ESR) data at room temperature). At the same time, the effective magnetic moment $\mu_{eff} \approx 6.5\mu_B$/f.u.for $Na_3Co_2SbO_6$ is slightly lower than the theoretical value using $g=3.3\pm0.1$ obtained from ESR data.

Specific heat data in zero magnetic field are in a good agreement with the temperature dependences of the magnetic susceptibility in weak magnetic fields, indicating the establishment of long-range magnetic order at low temperatures (Fig. 5). The temperature dependence $C_p(T)$ for $Na_3Co_2SbO_6$ (Fig. 5a) exhibit a distinct λ-anomaly at $T_N \sim 6.7$ K, while anomaly for $Li_3Co_2SbO_6$ centred at $T_N \sim 10$ K is rather weak and broad, which may possibly relate to a significant contribution of short-range order correlations in this compound (Fig. 5b).

For quantitative estimations, the specific heat data were measured also for nonmagnetic isostructural[45] analogue $Li_3Zn_2SbO_6$, assuming that the specific heat of the isostructural compound provides an estimate for the pure lattice contribution to the specific heat. The values for Debye temperature $\Theta_D$ have been estimated as $\Theta_D^{nonmag} = 515 \pm 5K$ for the diamagnetic compound $Li_3Zn_2SbO_6$; $\Theta_D^{mag} = 645 \pm 5K$ and $\Theta_D^{mag} = 507 \pm 5K$ for $Na_3Co_2SbO_6$ and $Li_3Co_2SbO_6$, respectively, taking into account the difference between the molar masses for Li – Na and Zn – Co atoms.[46] It's established that the specific heat jump at Néel temperature amounts $\Delta C_p \approx 7$ and $3.7$ J/(mol K) for $Na_3Co_2SbO_6$ and $Li_3Co_2SbO_6$, respectively (insets on Fig. 5), what is much less than the values predicted from the mean-field theory for the antiferromagnetic spin ordering assuming cobalt to be in high-spin state $Co^{2+}$ ($S=3/2$) $\Delta C_p = 5R \cdot S(S+1)/[S^2+(S+1)^2] \approx 36.7$ J/(molK).[46] This indicates the presence of appreciable short-range correlations far above $T_N$, which is typical for frustrated and LD systems.[47,48] We analyse the magnetic counterpart to the specific heat $C_m(T)$ below $T_N$ in terms of the spin-wave (SW) approach assumingthe limiting low-temperature behaviour of the magnetic specific heat should follow $C_m \propto T^{d/n}$ power law for magnons,[48] where $d$ stands for the dimensionality of the magnetic latticeand$n$ is defined as the exponent in the dispersion relation $\omega \approx \kappa^n$.

For $Na_3Co_2SbO_6$ the least-squares fitting of the data below $T_N$ (inset in Fig. 5a) has given with good accuracy $d = 3$ and $n = 1$ values, that implies presence of 3D AFM magnons at the lowest temperatures. However, in case of $Li_3Co_2SbO_6$, similar fitting procedure has given $d = 2$ and $n = 1$ values (inset in Fig. 5b), confirming picture of the 2D AFM magnons in this compound.

It was established that the magnetic entropy $\Delta S_m$ saturates at about 40 K reaching approximately $\Delta S_m \approx$ 7.5 and 5 J/(mol K) for $Na_3Co_2SbO_6$ and $Li_3Co_2SbO_6$, which is significantly less (only ~ 30% and 20%) to those expected from the mean-field theory for spin system of two $Co^{2+}$ ions with $S=3/2$: $\Delta S_m=2R\ln(2S+1)\approx23$ J/(mol K).[46] Below $T_N$ the magnetic entropy released removes only about 3J/(mol K) ( ~13%) and 2 J/(mol K) (less than 10%) of the saturation value for $Na_3Co_2SbO_6$ and $Li_3Co_2SbO_6$ respectively. This reveals the existence of noticeable frustration in both compound and the short-range correlations far above $T_N$, which is characteristic feature of 2D magnetism for both systems under study.[47]

The field dependence of the magnetization $M(B)$ for $A_3Co_2SbO_6$ (A = Na, Li) samples almost reach a saturation in moderate magnetic fields up to 9T (Fig. 6). The magnetization isotherms $M(B)$ do not display hysteresis, but indicate a magnetic field induced spin-reorientation (spin flop type) transitions. Moreover, a careful analysis the derivative $dM/dB(B)$ for both sodium and lithium antimonates shows the presence of two pronounced maxima, possibly corresponding to two successive spin-reorientation transitions at low temperatures. With increasing temperature, the positions of these anomalies shift slightly towards the lower fields and become almost undetectable above the Neel temperature. The critical fields BC were estimated from the magnetization derivative $dM/dB$: $B_{C1}$ = 0.5T, $B_{C2}$ = 1.2T for $Na_3Co_2SbO_6$, and $B_{C1}$ = 0.2T, $B_{C2}$ = 0.6T for $Li_3Co_2SbO_6$, respectively.

### 3.3. Low-temperature neutron powder diffraction

The results of low-temperature neutron powder diffraction at G4.1 diffractometer are presented on Fig.7a. The intensity of additional reflections increases with decreasing temperature and drops sharply with the scattering angle. The appearance of a set of new peaks, especially before the first nuclear peak (001) position, corresponds to the long-range antiferromagnetic ordering around 11 K. Note that, according to the neutron diffraction data, the Néel temperature is slightly higher than obtained from the magnetic susceptibility measurements but is in agreement with the specific heat data. At $T$ = 11 K, a weak magnetic satellite peak near the $2\theta = 14°$ is still observed, which has the maximum intensity among the magnetic reflections at $T$ = 1.5 K.

In order to determine magnetic structure, we used the difference powder pattern between 1.5 K and 30 K, which corresponds to the "pure" magnetic contribution to the neutron scattering (Fig. 7b). All the magnetic peaks can be indexed with the commensurate propagation vector **k** = (0 0 ½) that indicates the doubled magnetic cell along the $c$-crystallographic axis in comparison with the crystal lattice. Since there is only one magnetic atom position, it automatically leads to the antiferromagnetic interaction between the adjacent honeycomb layers.

In order to get information about the magnetic lattice symmetry, we have performed representation analysis with BASIREPS program implemented in Fullprof Suite. The magnetic representation $\Gamma_{mag}$ based on the space group $C2/m$ and the propagation vector **k** = (0 0 ½) consists of four one-dimensional irreducible representations (IRs). The basis vectors of these IRs (i.e., the Fourier components of the magnetization) are given in Table 5.

The first and the second IRs have one basis vector, which allows only nonzero magnetic moment component along the b-crystallographic direction, whereas the other two lead to the magnetic moments within ac-plane. Besides, the magnetic ordering in the honeycomb layers depends on the symmetry transformations: $\Gamma_1$ and $\Gamma_3$ correspond to the Neel magnetic state with antiferromagnetic nearest neighbour alignment while $\Gamma_2$ and $\Gamma_4$ correspond to the ferromagnetic state.

All irreducible representations were tested in order to fit neutron diffraction data. For a new magnetic phase, we used a scale factor, the peak shape parameters, and the crystal structure parameters determined from the Rietveld refinement of the neutron powder pattern at $T$ = 30 K. It was found, that the Neel magnetic ordering strongly disagrees the experimental data, whereas the ferromagnetic ordering provides an acceptable result. The best fitting obtained for $\Gamma_1$ irreducible representation is presented on Fig. 7b. It should be also noted that a similar result was observed for $\Gamma_3$ irreducible representation but with worse R- and $\chi^2$- factors. The corresponding magnetic structure is shown on Fig. 8. This is represented by ferromagnetic arrangement within the honeycomb layers with the antiferromagnetic coupling between the adjacent layers.

Thus, our results demonstrate that $Li_3Co_2SbO_6$ has magnetic ordering which is different from the previously studied related compounds. Most of them tend to a zigzag AFM structure that has been repeatedly reported for complex oxides with honeycomb structure. While the resulting magnetic ground state depends on the magnetic interactions on the honeycomb net, the ferromagnetic ordering in $Li_3Co_2SbO_6$ can be caused by strong coupling of nearest $Co^{2+}$ ions with relatively large spin S = 3/2. Nevertheless, in our case, cobalt magnetic moments preserve the general trend to have in-plane anisotropy whereas $Ni^{2+}$ ions have out-plane direction. Based on the neutron powder diffraction data, the refined ordered magnetic moments take the value M = 3.74(4) $\mu_B$ at $T$ = 1.5 K. The observed magnitude is significantly smaller than theoretically predicted for $Co^{2+}$ (S=3/2). This is typical for diffraction experiment because neutrons can detect only the average magnetic moment, which can be reduced by the thermal fluctuations presence and frustration of the magnetic interactions. On the other hand, it may be associated with the stacking faults.

The observed magnetic structure in $Li_3Co_2SbO_6$ with FM honeycomb layers is in a complete accordance with the magnetic susceptibility data, which implied predominance of the ferromagnetic interactions in this compound. Contrarily, $Na_3Co_2SbO_6$ demonstrates zigzag magnetic ordering, where AFM couplings within the layers play an important role. Apparently, such difference is associated with the different alkali metal ions, which determine the interlayer distance and affect the magnetic interactions between the adjacent layers.

## 4. Conclusions

Monoclinic rather than trigonal symmetry of $Na_3Co_2SbO_6$ was demonstrated directly for the first time by peak splitting in the synchrotron powder pattern.Honeycomb-layered $Li_3Co_2SbO_6$ could only be prepared by Na/Li exchange in molten LiCl-LiBr mixture because ion-exchange in oxidizing medium ($LiNO_3$) or direct solid-state synthesis resulted in different phases. Rietveld refinement of Li$_3$Co$_2$SbO$_6$ within space group *C2/m* confirmed the O3-type superlattice with honeycomb ordering of Co and Sb and showed partial Li/Co site inversion. Studies of the thermodynamic properties have shown that both compounds order antiferromagnetically at low temperatures. At the same time, the competing ferro- and antiferromagnetic exchange interactions present over the wide temperature range that provide conditions to high sensitivity to the external magnetic fields. According to the neutron diffraction data from Li$_3$Co$_2$SbO$_6$, we found a type of magnetic ordering in the honeycomb compounds characterized by purely ferromagnetic ordering in a layer and antiferromagnetic ordering between layers. According to our knowledge, such a magnetic order is experimentally observed for the first time, at least in antimonates and tellurates with a layered honeycomb structure.

**References**

1. J.M.S. Skakle, M.A. Castellanos R., S. Trujillo Tovar, A.R. West, *J. Solid State Chem.*, 1997, **131**, 115-120.
2. O.A. Smirnova, V.B. Nalbandyan, V.V. Politaev, L.I. Medvedeva, V.A. Volochaev, I.L. Shukaev, B.S. Medvedev, A.A. Petrenko, *Solid State Chemistry 2000*, P. Bezdicka and T. Grygar (Ed.), Academy of Sciences of the Czech Republic, Prague, 2000, p. 228.
3. R. Nagarajan, S. Uma, M.K. Jayaraj, J. Tate, A.W. Sleight, *Solid State Sci.,* 2002, **4**, 787-792.
4. O.A. Smirnova, V.B. Nalbandyan, A.A. Petrenko, M. Avdeev, *J. Solid State Chem.*, 2005, **178**, 1165-1170.
5. X. Ma, K. Kang, G. Ceder, Y.S. Meng, *J. Power Sources*, 2007, **173**, 550-555.
6. P. Cui, Z. Jia, L. Li, T. He, *J. Phys. Chem. Solids*, 2011, **72**, 899-903.
7. N. Twu, X. Li, A. Urban, M. Balasubramanian, J. Lee, L. Liu, G. Ceder, *Nano Lett.*, 2015, **15**, 596-602.
8. D. Yuan, L. Wu, Y. Cao, X. Ai, H. Yang, *Adv. Mater.*, 2014, **26**, 6301-6306.
9. J. Ma, S.-H.Bo, L.Wu, Y.Zhu, C.P.Grey, P.G. KhaLifah, *Chem. Mater.*, 2015, **27**, 2387-2399.
10. F. Aguesse, J.-M. Lopez del Amo, L. Otaegui, E. Goikolea, T. Rojo, G. Singh, *J. Power Sources*, 2016, **336**, 186-195.
11. Y. Kee, N. Dimov, A. Staykov, S. Okada, *Mater. Lett.*, 2016, **183**, 187 – 190.
12. Y. You, S.O. Kim, A. Manthiram, *Adv. Energy Mater*, 2016, **7**, 1601698.
13. D. Gyabeng, D.A.Anang, J.I.Han, *J. Alloys Comp.*, 2017, **704**, 734-741.
14. H. Dai, C. Yang, X. Ou, X. Liang, H. Xue, W. Wang, G. Xu, *Electrochim. Acta*, 2017, **257**, 146-154.
15. P.-F. Wang, H.-R. Yao, Y. You, Y.-G. Sun, Y.-X. Yin, Y.-G. Guo, *Nano Res.*, 2018, **11**, 3258–3271.
16. S.-Y. Xu, X.-Y. Wu, Y.-M. Li, Y.-S. Hu, L.-Q. Chen, *Phys. B*, 2014, **23**, 118202.
17. Y. Miura, R. Hirai, Y. Kobayashi, M. Sato, *J. Phys. Soc. Japan*, 2006, **75**, 084707.
18. Y. Miura, R. Hirai, T. Fujita, Y. Kobayashi, M. Sato, *J. Magn. & Magn. Mater*, 2007, **310**, e389-e391.

19. S. Derakhshan, H.L. Cuthbert, J.E. Greedan, B. Rahaman, T. Saha-Dasgupta, *Phys. Rev. B*, 2007, **76**, 104403.
20. Y. Miura, Y. Yasui, T. Moyoshi, M. Sato, K. Kakurai, *J. Phys. Soc. Japan*, 2008, **77**, 104709.
21. H.-J. Koo, M.-H. Whangbo, *Inorg. Chem.*, 2008, **47**, 128-133.
22. C.N. Kuo, T.S. Jian, C.S. Lue, *J. Alloys Comp.*, 2012, **531**, 1-4.
23. E. Climent-Pascual, P. Norby, N.H. Andersen, P.W. Stephens, H.W. Zandbergen, J. Larsen, R.J. Cava, *Inorg. Chem.*, 2012, **51**, 557.
24. M. Schmitt, O. Janson, S. Golbs, M. Schmidt, W. Schnelle, J. Richter, H. Rosner, *Phys. Rev. B.*, 2014, **89**, 174403.
25. C. Koo, E.A. Zvereva, I.L. Shukaev, M. Richter, M.I. Stratan, A.N. Vasiliev, V.B. Nalbandyan, R. Klingeler, *J. Phys. Soc. Japan*, 2016, **85**, 084702.
26. W. Schmidt, R. Berthelot, A.W. Sleight, M.A. Subramanian, *J. Solid State Chem.*, 2013, **201**, 178–185.
27. E.A. Zvereva, M.A. Evstigneeva, V.B. Nalbandyan, O.A. Savelieva, S.A. Ibragimov, O.S. Volkova, L.I. Medvedeva, A.N. Vasiliev, R. Klingeler, B. Büchner, *Dalton Trans.*, 2012, **41**, 572-580.
28. E.A. Zvereva, M.I. Stratan, Y.A. Ovchenkov, V.B. Nalbandyan, J.-Y. Lin, E.L. Vavilova, M.F. Iakovleva, M. Abdel-Hafiez, A.V. Silhanek, X.-J. Chen, A. Stroppa, S. Picozzi, H.O. Jeschke, R. Valentí, A.N. Vasiliev, *Phys. Rev. B,* 2015, **92**, 144401.
29. A.I. Kurbakov, A.I. Korshunov, S.Yu. Podchezertsev, A.L. Malyshev, M.A. Evstigneeva, F. Damay, J. Park, C. Koo, R. Klingeler, E.A. Zvereva, V.B. Nalbandyan, *Phys. Rev. B*, 2017, **96**, 024417.
30. J.H. Roudebush, G. Sahasrabudhe, S.L. Bergman, R.J. Cava, *Inorg. Chem.*, 2015, **54**, 3203-3210.
31. L. Viciu, Q. Huang, E. Morosan, H.W. Zandbergen, N.I. Greenbaum, T. McQueen, R.J. Cava, *J. Solid State Chem.*, 2007, **180**, 1060-1067.
32. C. Wong, M. Avdeev, C.G. Ling, *J. Solid State Chem.*, 2016, **243**, 18–22.
33. E.A. Zvereva, M.I. Stratan, A.V. Ushakov, V.B. Nalbandyan, I.L. Shukaev, A.V. Silhanek, M. Abdel-Hafiez, S.V. Streltsov, A.N. Vasiliev, *Dalton Trans.,* 2016, **45**, 7373 – 7384.
34. A.S. Ermolov, E.A. Zvereva, M.I. Stratan, V.B. Nalbandyan, M.A. Evstigneeva, A.N. Vasiliev, *Proceedings of XVI International Youth Scientific School "Actual problems of magnetic resonance and its application"*, Kazan: Kazan Federal University, 2013, 40-42.
35. Powder Diffraction File. *International Centre for Diffraction Data*, Newtown Square, Pennsylvania, USA. 2008. Cards 00-58-637 and 00-58-770.
36. G.C. Mather, R.I. Smith, J.M.S. Skakle, J.G. Fletcher, M.A. Castellanos R, M.P. Gutierrez, A.R. West, *J. Mater. Chem.*, 1995, **5**, 1177-1182.
37. Powder Diffraction File. *International Centre for Diffraction Data*, Newtown Square, Pennsylvania, USA. 2008. Card 00-58-771.
38. V.V. Politaev, V.B. Nalbandyan, A.A. Petrenko, I.L. Shukaev, V.A. Volotchaev, B.S. Medvedev. *J. Solid State Chem.*, 2010, **183**, 684-691.
39. R.D. Shannon, *Acta Crystallogr. A*, 1976, **32**, 751-767.
40. A.C. Larson, R.B. Von Dreele, *LAUR 86-748*, Los Alamos National Laboratory, Los Alamos, NM, 2004.

41 B.H. Toby, *J. Appl. Crystallogr.*, 2001, **34**, 210.
42 J. Rodríguez-Carvajal, *Physica B*, 1993, **192**, 55
43 FULLPROF suite, http://www.ill.eu/sites/fullprof/.
44 O.C. Gagné, F.C. Hawthorne, *Acta Crystallogr. B*, 2015, **71**, 562-578.
45 C. Greaves, S.M.A. Katib, *Mat. Res. Bull.,* 1990, **25**, 1175-1182.
46 A. Tari, *The Specific Heat of Matter at Low Temperature*, Imperial College Press, London, 2003.
47 R.L. Carlin, *Magnetochemistry*; Springer-Verlag, Berlin, 1986.
48 L.J. deJongh, A.R. Miedema, *Adv. Phys.*, 1974, **23**, 1–260.

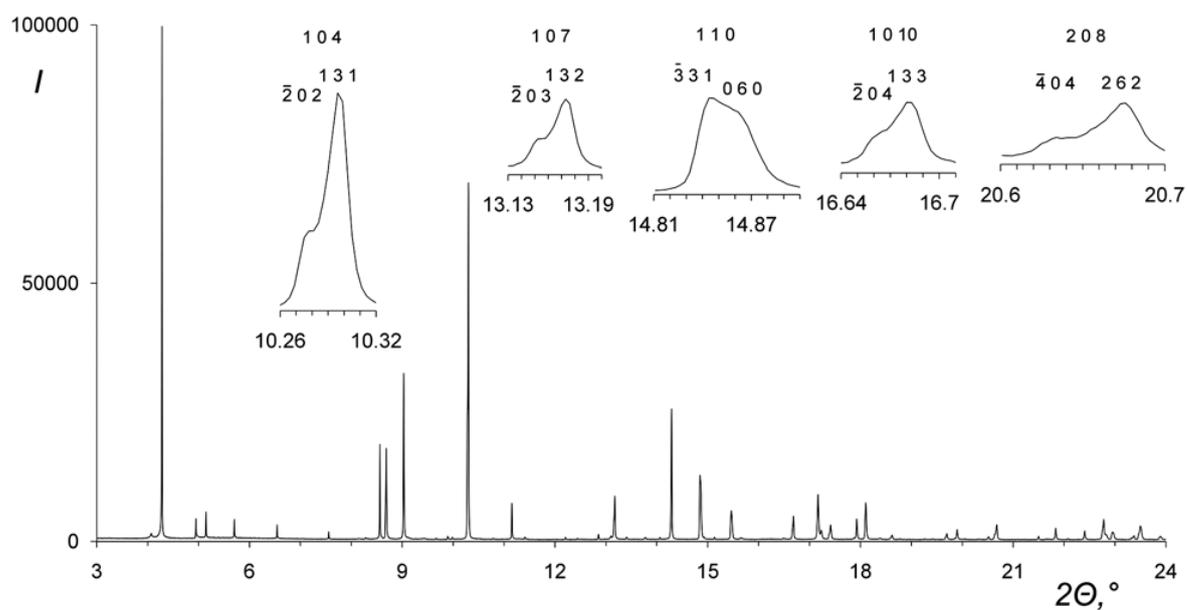

**Fig. 1**. Synchrotron diffraction pattern of $Na_3Co_2SbO_6$. Insets illustrate splitting of certain pseudohexagonal reflections. Upper row: hkl referred to the α-$NaFeO_2$-type subcell ($R\bar{3}m$) in hexagonal setting. Lower row: hkl referred to the monoclinic (C2/m) unit cell.

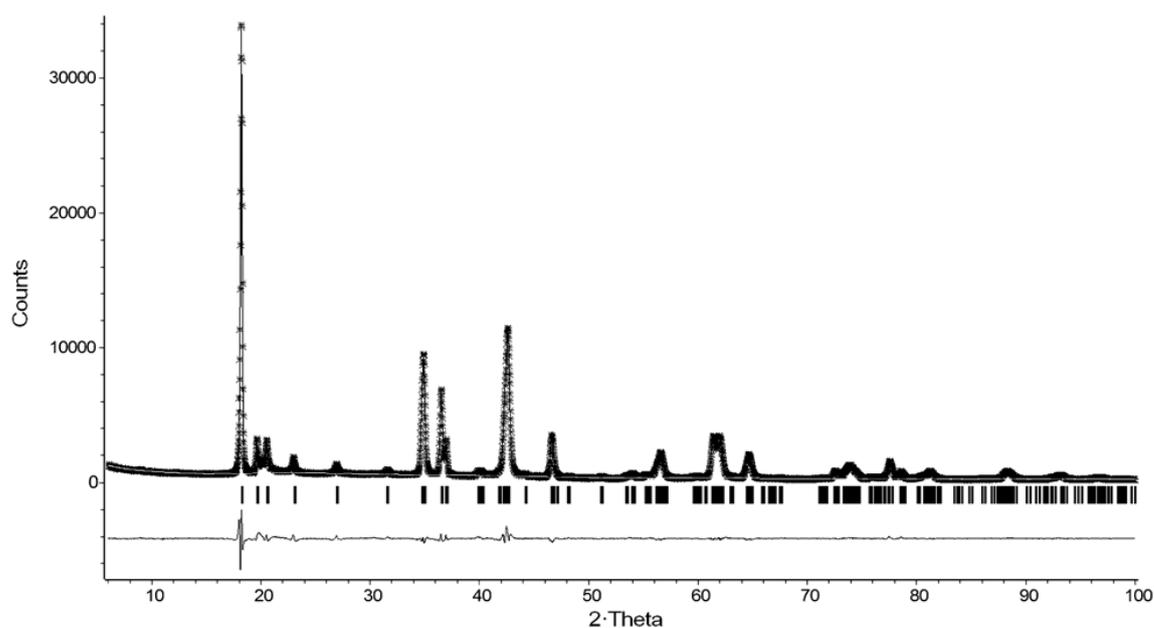

**Fig. 2.** XRD profile of $Li_3Co_2SbO_6$ after structure refinement. Asterisks, experimental points; solid line, calculated profile; thin bottom line, difference profile; vertical bars, calculated positions of Bragg reflection

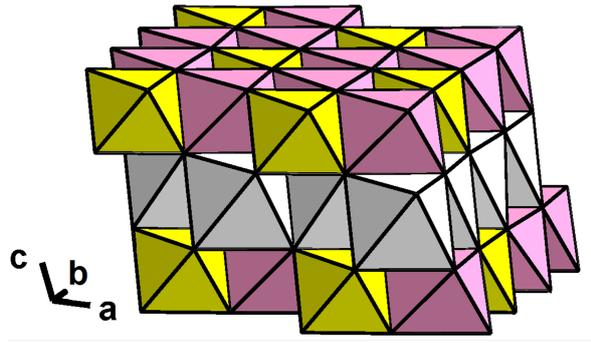

**Fig. 3.** Crystal structure of $Li_3Co_2SbO_6$ in polyhedral presentation. Yellow octahedra, $SbO_6$; pink octahedra, $CoO_6$ (with small Li substitution); grey octahedra, $LiO_6$ (with small Co substitution)

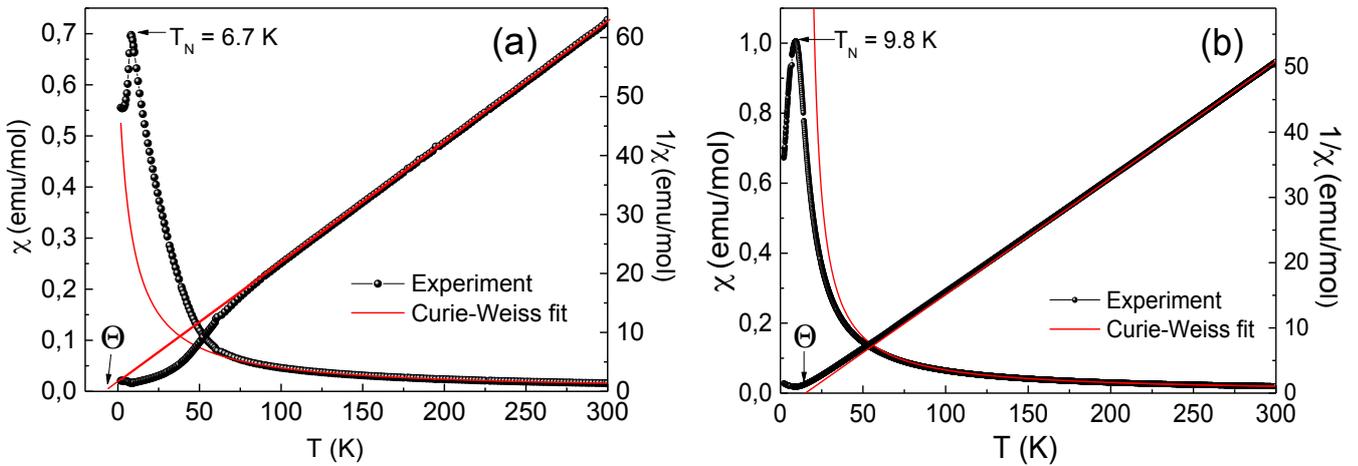

**Fig. 4.** The temperature dependences of the static magnetic susceptibility $\chi(T) = M/B$ with their inverse values $1/\chi$ at $B = 0.1$ T for $Na_3Co_2SbO_6$ (a) and $Li_3Co_2SbO_6$ (b). The solid red lines represent the Curie-Weiss fit of the high temperature data.

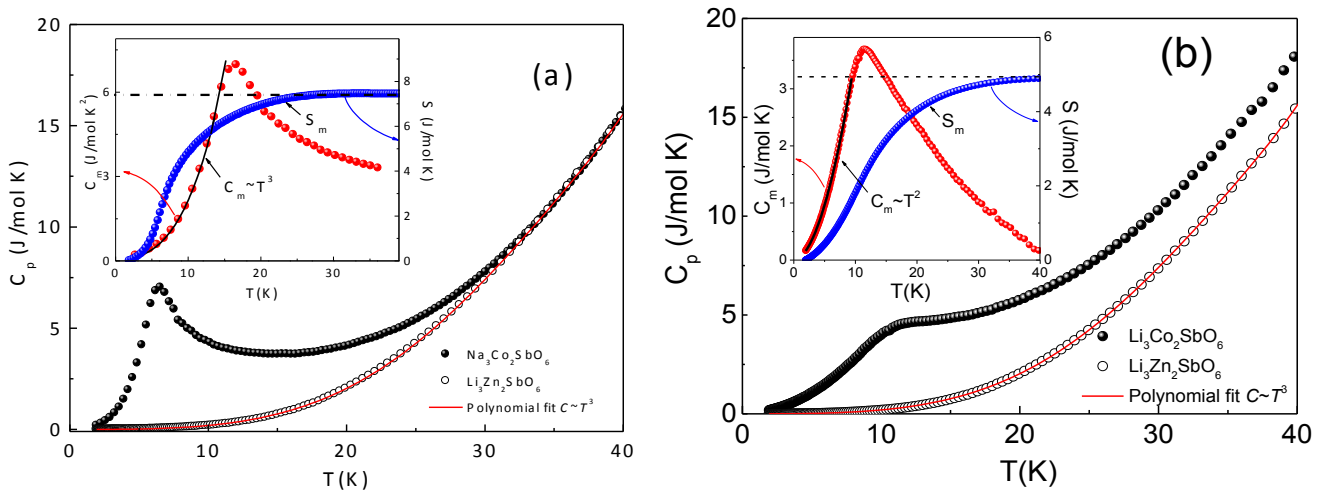

**Fig. 5.** Temperature dependences of the specific heat in $Na_3Co_2SbO_6$ (a) and $Li_3Co_2SbO_6$ (b) (filled symbols) and non-magnetic isostructural analogue $Li_3Zn_2SbO_6$ (open circles) in zero magnetic field. Red lines represent Debye approximation. On the insets: temperature dependences of $C_m(T)$ (red symbols) and the entropy $S_m(T)$ (blue symbols). The black line is the result of approximation in the framework of the spin waves theory.

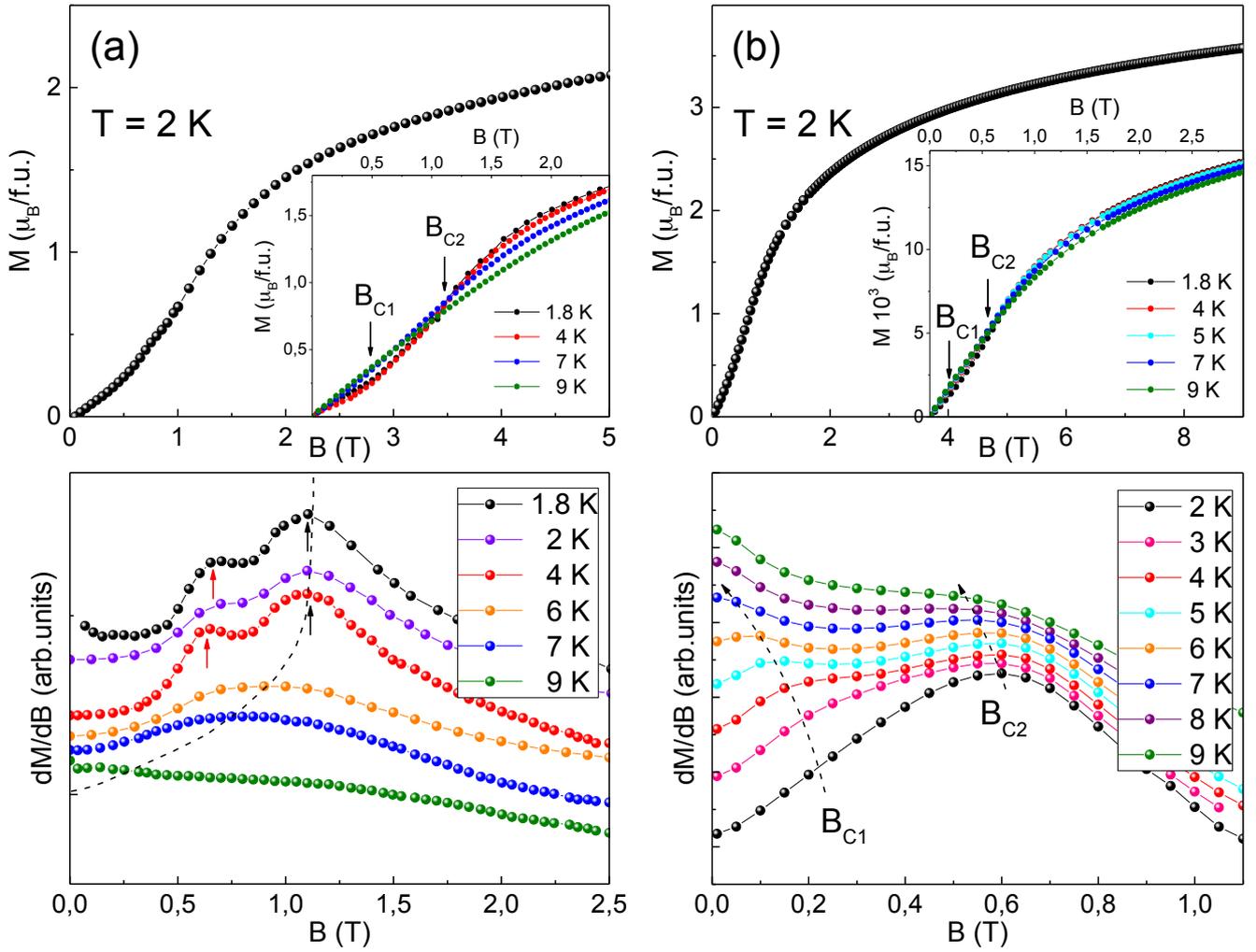

**Fig. 6.** On upper panels: the magnetization isotherms at $T$=2K for $Na_3Co_2SbO_6$ (a) and $Li_3Co_2SbO_6$(b); on insets: the magnetization curves for both compounds at various temperatures. On lower panels: the magnetization derivative d$M$/d$B$ for $(Na,Li)_3Co_2SbO_6$ at various temperatures. The dashed lines and arrows indicate the position of spin-reorientation transition, observed at $T<T_N$.

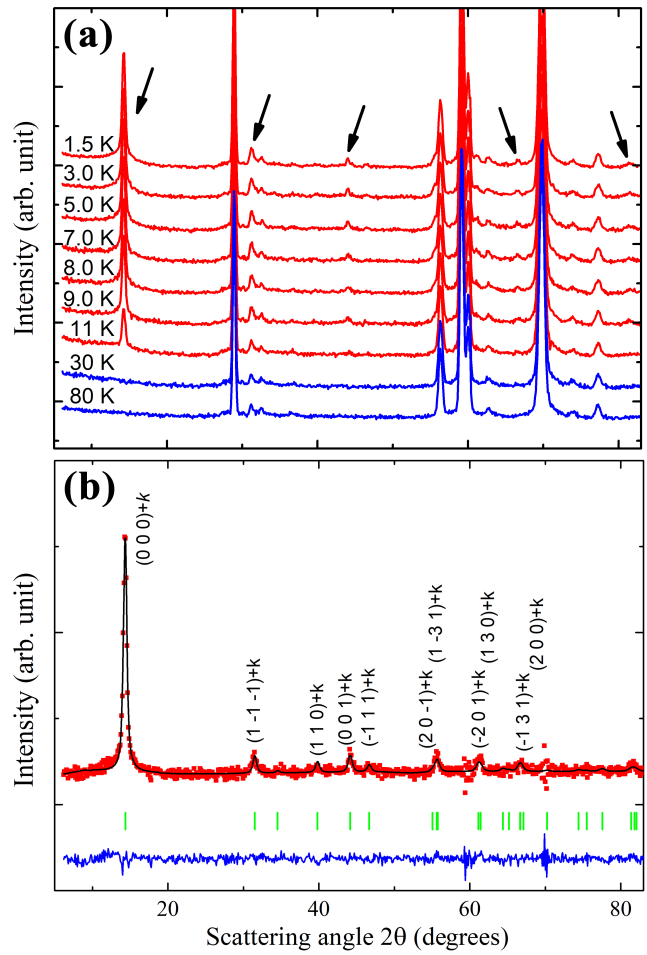

**Fig.7.** (a) Experimentally measured neutron diffraction patterns (G4.1, λ = 2.428Å) in temperature range 1.5K – 80K. Arrows are pointing to the most intense additional reflections associated with the long-range magnetic ordering in $Li_3Co_2SbO_6$. (b) The refined magnetic diffraction pattern with k = (1/2 0 0) without nuclear contribution at 1.5 K. The difference between observed (red dots) and calculated (black line) patterns are shown by the solid blue line at the bottom. The vertical bars indicate the positions of magnetic Bragg peaks.

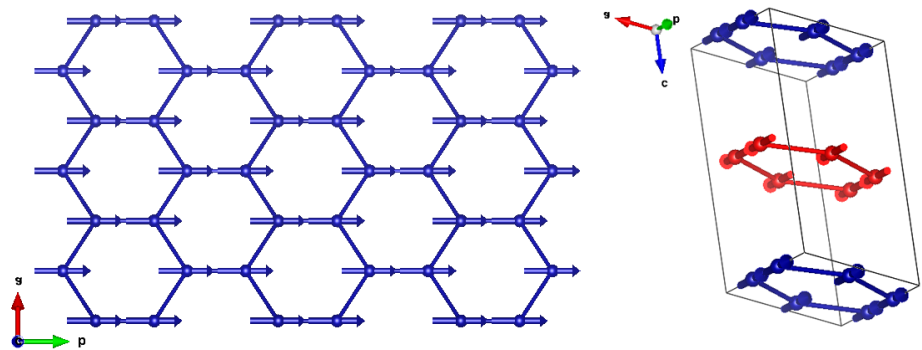

**Fig.8.** The magnetic structure of $Li_3Co_2SbO_6$ at 1.5 K corresponded $\Gamma_1$ irreducible representation. Different colours denote opposite directions of magnetic moments.

**Table 1.** Comparison of Na$_3$Co$_2$SbO$_6$ lattice parameters from the four powder diffraction studies

| Method | $a$, Å | $b$, Å | $c$, Å | $\beta$, ° |
|---|---|---|---|---|
| Laboratory XRD, R-3m $R\bar{3}m$ subcell[37,38] | 3.0932(4) | | 16.0733(11) | |
| Same, monoclinic indexing[38] | 5.3627(21) | 9.2790(11) | 5.6514(18) | 108.55(4) |
| Neutron diffraction[31] | 5.3681(2) | 9.2849(4) | 5.6537(2) | 108.506(4) |
| Neutron diffraction[32] | 5.35648(10) | 9.28723(19) | 5.65100(7) | 108.358(1) |
| Synchrotron XRD (this work) | 5.3649(6) | 9.2782(3) | 5.6533(5) | 108.490(8) |

**Table 2.** Details of the data collection and structure refinement of Li$_3$Co$_2$SbO$_6$

| Crystal system | | Monoclinic | Density (calc.) | | 5.133 |
|---|---|---|---|---|---|
| Space group | | C2/m (no. 12) | Texture parameters (March-Dollase) | | axis 001 ratio 0.993 |
| Lattice constants | $a$, Å | 5.24111(16) | 2Θ range, º | | 6.02–99.98 |
| | $b$, Å | 9.03910(23) | 2Θ step, º | | 0.02 |
| | $c$, Å | 5.18179(12) | Anisotropic broadening | | Axis 001 |
| | $\beta$, ° | 110.0493(22) | Stacking faults sublattice vectors | | 210 and 030 |
| Cell volume, Å$^3$ | | 230.610(11) | No. of data points | | 4699 |
| Formula weight | | 356.43 | No. of reflections calc. ($\alpha_1$ only) | | 127 |
| Z | | 2 | No. of variables | | 56 (14 – of structure) |
| Wavelengths, Å | $\alpha_1$ | 1.54056 | Agreement factors | GOF | 2.25 |
| | $\alpha_2$ | 1.54439 | | R(F$^2$) | 0.03259 |
| | Ratio | 0.5 | | R$_p$ | 0.0558 |
| Polarization ratio | | 0.707 | | R$_{wp}$ | 0.0753 |
| | | | | $\chi^2$ | 5.067 |

**Table 3.** Atomic positions, occupancies and thermal displacement parameters for $Li_3Co_2SbO_6$

| Site | Wyckoff position | Site symmetry | Atom | Occupancy | x/a | y/b | z/c | $U_{iso}$ |
|------|------------------|---------------|------|-----------|-----|-----|-----|-----------|
| Sb   | 2a  | 2/m | Sb | 1 | 0 | 0 | 0 | 0.00219(2) |
| Co   | 4g  | 2   | Co<br>Li | 0.95**<br>0.05* | 0 | 0.3346(3) | 0 | 0.0206(4) |
| Li1  | 4h  | 2   | Li<br>Co | 0.958(2)<br>0.042** | 0 | 0.1653(20) | 1/2 | 0.02* |
| Li2  | 2d  | 2/m | Li<br>Co | 0.985(4)<br>0.015** | 0 | 1/2 | 1/2 | 0.02* |
| O1   | 4i  | m   | O | 1 | 0.7643(15) | 0 | 0.2336(12) | 0.02* |
| O2   | 8c  | 1   |   | 1 | 0.2287(11) | 0.1596(6) | 0.2256(7) | 0.02* |

**Table 4.** Bond lengths (Å) and bond valence sums in $Li_3Co_2SbO_6$

| Bonds | Sb-O | Co-O | Li1-O | Li2-O |
|-------|------|------|-------|-------|
| Distances | 1.981(5)×4<br>2.004(8)×2 | 2.084(5)×2<br>2.114(5)×2<br>2.129(6)×2 | 2.120(14)×2<br>2.152(5)×2<br>2.272(14)×2 | 2.177(4)×4<br>2.266(8)×2 |
| Average | 1.99 | 2.11 | 2.18 | 2.21 |
| Sum of ionic radii[39] | 2.00 | 2.15 | 2.16 | 2.16 |

| Atom | Sb | Co | Li1 | Li2 | O1 | O2 |
|------|-----|-----|-----|-----|-----|-----|
| Bond valence sum | 4.90 | 2.01 | 1.05 | 1.01 | 1.99 | 2.02 |

**Table 5.** Basis vectors of irreducible representations for the propagation vector k = (0 0 1/2) and the space group C2/m. The symmetry operators are written for magnetic Co2+ in 4g Wyckoff position with coordinates (0, 0.334, 0).

| IRs | | Basis Vectors | | Magnetic ground state |
|-----|---|---------------|---|------------------------|
|     |   | x,y,z | -x,-y+1,-z | |
| $\Gamma_1$ | $\Psi_1$ | (010) | (010) | FM |
| $\Gamma_2$ | $\Psi_1$ | (010) | (0−10) | Neel |
| $\Gamma_3$ | $\Psi_1$ | (100) | (100) | FM |
|            | $\Psi_2$ | (001) | (001) | |
| $\Gamma_4$ | $\Psi_1$ | (100) | (−100) | Neel |
|            | $\Psi_2$ | (001) | (00−1) | |